\documentstyle[11pt,epsf,twoside]{article}

\begin{document}
\title{Comments On String Theory}

\vskip 1cm
\author{Edward Witten\thanks{Research supported in part by NSF Grant
PHY-0070928.  This manuscript is based on a lecture presented at the Sackler
Colloquium on Challenges to the Standard Paradigm (National Academy of Sciences,
October 2002).} \\
Institute for Advanced Study\\ Princeton, NJ 08540 USA}
\vskip 1cm

\maketitle 

\begin{abstract}
String theory avoids the ultraviolet infinities that arise in trying to quantize
gravity. It is also more predictive than conventional quantum field theory,
one aspect of this being the way that it   contributed to the emergence of the concept 
of ``supersymmetry'' of particle interactions.  There are hints from the successes of
supersymmetric unified theories of particle interactions that supersymmetry
is relevant to elementary particles at energies close to current accelerator energies;
if this is so, it will be confirmed experimentally
and supersymmetry is then also likely to be important in cosmology, in connection
with dark matter, baryogenesis, and/or inflation.   Magnetic monopoles play an important
role in the structure of string theory, and thus should certainly exist, if string theory
is correct, though they may have been diluted by inflation to an unobservable level.  The monopole
mass in many attractive models is near the Planck mass, but, if unification
of elementary particle forces with gravity occurs near TeV energies through large or warped
extra dimensions, as in some recent models, then monopoles should be below 100 TeV and in an 
astrophysical context would be ultrarelativistic.  In such models, supersymmetry would 
definitely be expected at TeV energies.  
\end{abstract}
\vskip 2.5cm

In string theory, at the most basic level, an elementary particle
is treated as a vibrating string, rather than a point particle.
A string has many different harmonics of vibration, and in this context
different elementary particles are interpreted as different harmonics
of the string.

In a Feynman diagram of ordinary quantum field theory, there are interaction
vertices at which particles split and rejoin (figure 1(a)).  These are definite spacetime
events that every Lorentz observer can agree upon.  By contrast,
in the splitting and rejoining of a string (figure 1(b)), there is no
definite moment at which an interaction occurred.  If one looks at the whole
history, it is clear that one string split into two, but nevertheless
any small piece of the history looks like any other.  This is a hint
of one of the most important facts about the theory:
once one decides what the string is, the interactions follow; they need
not be separately postulated, as is the case in quantum field theory.

Quantizing a relativistic string proves to be very subtle.  When
this was done in the 1970's, some highly unusual features appeared.  Perhaps
the most dramatic is that the ground state of the (closed)
string proves to be a massless spin two field, i.e. a graviton.  The 
interactions do mimic those of general relativity at long wavelengths,
so  string theory proves to be a quantum theory of gravity.
This is something that we need in physics -- since both quantum mechanics and 
gravity are present in nature -- and and it is something that we otherwise do not have,
since conventional attempts at quantizing general relativity are thwarted by ultraviolet
divergences.

The ultraviolet divergences of quantum general relativity
disappear when one replaces a particle by a string.   This is so roughly
because of the fact that was illustrated in figure 1: replacing particles
by strings has the effect of smearing out the interaction vertex in 
a Feynman diagram.  



\begin{figure}
\begin{center}
\leavevmode
\epsfysize=2in
\epsffile{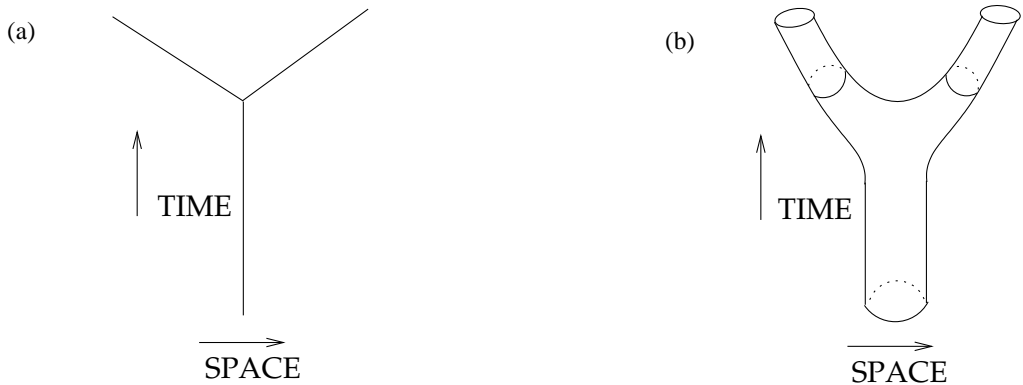}
\end{center}
\caption{(a) A space-time history showing one particle breaking into two. (b) An analogous
history in string theory.}
\label{fig:space-time}
\end{figure}

String theory also leads to {\it gauge symmetry} in much the same way that
it leads to gravity.  For example, the ground state of an open string
 turns out to be a gauge field.  (A closed string is a loop of string that closes
 on itself without ends; an open string does have ends.)

\bigskip\noindent{\it Supersymmetry}

Along with gauge theory and gravity, string theory generates
``supersymmetry,'' which is symmetry between bosons and fermions.
In fact, the appearance around 1970
of what we now call worldsheet supersymmetry in the Ramond model of
string theory was part of the historical origin of the idea of supersymmetry.
Nature does not have exact supersymmetry.  But it is possible that
nature does have an underlying supersymmetry which is ``spontaneously broken,''
like the $SU(2)\times U(1)$ gauge symmetry of the standard model
of particle physics.  In fact, it may be that supersymmetry is
detectable at energies relevant to current and planned accelerator
experiments.  There are a few hints of this.  

One is the ``hierarchy 
problem,'' which is a modern version of Dirac's ``problem of large numbers.''
In Dirac's formulation, why is the gravitational force between two protons
$10^{-38}$ times weaker than the electric force?  The appearance of
such a tiny dimensionless number in the laws of physics seems to require
some explanation.  An updated version of the question is to ask why the
$W$ and $Z$ bosons (the gauge bosons whose masses are ultimately linked
to the mass scale of other particles) have masses $10^{-17}$
times smaller than the Planck mass.  At any rate, supersymmetry offers
a possible solution to this problem, because supersymmetry removes the
quadratic divergences that otherwise affect the Higgs boson mass.

A more quantitative hint of supersymmetry comes 
from the measured values of the strong, weak, and electromagnetic coupling
constants.  They agree at about the $1\%$ level with a relation that
follows from supersymmetry together with grand unification of the
elementary particle forces.

If supersymmetry is discovered (presumably at Fermilab or at the LHC, the new
accelerator that is being built at CERN), 
there is much to be learned about the masses and couplings of the 
supersymmetric particles.  Theoretical models of the details of the
supersymmetric world are currently all over the map.  Even if one of the them is on
the right track, we have no idea which one.

Discovering supersymmetry would undoubtedly give string theory a big
boost, showing that the three structures -- gravity, gauge theory, and supersymmetry --
that arise from string theory
in roughly the same way are
all part of the description of nature.  It is impossible right now
to guess how big a boost string theory might get from the discovery and
exploration of supersymmetry, because we do not know what the pattern
of supersymmetric masses would turn out to be and therefore what clues we
would get about physics at still higher energies.

There are also several possibilities for how discovering supersymmetry
might be relevant to cosmology:

(1) Some of the supersymmetric particles make plausible dark matter
candidates, as calculations show that they would have just about the
right mass and abundance to agree with observation.  (However, supersymmetric
particles are not the only plausible particle physics candidates for
dark matter, and some supersymmetric models do not have such candidates.)
 
(2) If supersymmetry is correct, this must be taken into account
in theories of baryogenesis in the early universe.  In fact,
supersymmetric theories have scalar fields with baryon number
that might very well have played an important role \cite{dine}.

(3)  Supersymmetric scalar fields might also be relevant to inflation
(this was discussed in L. Randall's talk), though supersymmetry and
strings have not given a clear picture yet.

In 1984, string-based models of particle physics became much more 
interesting when Green-Schwarz anomaly cancellation and the construction of
the heterotic string by Gross, Harvey, Martinec, and Rohm made possible
models of particle physics plus quantum gravity that are elegant and
semi-realistic.  In this context, ``semirealistic'' means that one
neatly gets the right particles and gauge forces, but one does not have
a reasonable picture of particle masses, because these involve supersymmetry
breaking, for which we have no sensible model.

A good model of supersymmetry breaking ought to shed light on the 
extreme smallness (or vanishing?) of the cosmological constant, since
in our semi-realistic models, the cosmological constant vanishes when
supersymmetry is unbroken. 
Thus, not only is the extreme smallness of the cosmological constant 
a big mystery -- sharpened by observations suggesting that it is not zero --
but not understanding the cosmological constant makes it hard to improve
the models of particle physics.

Supersymmetry breaking models that we have now lead to 
quintessence-like behavior (an evolving scalar field and no 
stable vacuum) but with highly unrealistic parameters and 
couplings. In general, quintessence with a scalar field seems 
troublesome due to coherent couplings that would likely already 
have been detected, notably in tests of the equivalence 
principle.  In that respect, quintessence with an evolving 
pseudo-scalar (an axion-like field with a potential like 
$V(a)=\Lambda^4(1- \cos(a/F))$ with some constants $\Lambda$ and $F$) seems 
more attractive, given the absence (or extreme suppression, given 
that parity is not an exact symmetry of nature) of coherent 
couplings for pseudoscalars. There have only been a few papers 
on quintessence-like models based on pseudoscalars \cite{carroll}, \cite{choi}.

\bigskip\noindent{\it Five String Theories}

By 1984, there were five string theories, differing by very general 
properties of the string:

(1) In Type IIA and Type IIB, the strings are closed, oriented,
and insulating.

(2) Type I strings are open or closed and insulating; open Type I strings have
electric charges at the ends.  (This model has an analogy with strong
interaction physics, with the open and closed strings corresponding
respectively to mesons and glueballs; the analogy played a role in the
discovery of string theory and continues to motivate much current research.)

(3) Finally, heterotic $SO(32)$ and $E_8\times E_8$ strings are closed,
oriented, and superconducting.

Five string theories, each of which includes gravity,
 is four too many, but it is much better than the
situation in pre-string physics, where there are infinitely many possible
quantum field theories, none of which include gravity.

Here is another characteristic difference between string theory and 
pre-string physics.  In quantum field theory, we generally have adjustable
dimensionless parameters, such as the fine structure constant
$e^2/4\pi \hbar c\sim 1/137$ or the electron-muon mass ratio $m_e/m_\mu$, 
which is about
$1/200$.  These adjustable parameters mostly appear at the interaction
vertices of figure 1, and they disappear in going to string theory.
In string theory, there are no adjustable dimensionless parameters, but
instead there are scalar fields $\phi_i$ whose expectation values
determine $e^2/4\pi \hbar c$, $m_e/m_\mu$, etc. 

This means that in principle $e^2/4\pi \hbar c$, $m_e/m_\mu$, and the rest
might be computed by minimizing the energy as a function of the $\phi$'s.
In practice, to do this we would have to understand supersymmetry
breaking and the cosmological constant, since in the absence of
supersymmetry breaking, $V(\phi)$ is identically zero.  This is another
reason that it would be good to discover supersymmetry at accelerators
and thus have the chance to explore supersymmetry breaking experimentally.

\bigskip\noindent{\it Strong Coupling}

In trying to answer any of the really big questions in string theory,
we run against the fact that we do not really understand what the theory is.
In contrast to (say) general relativity, where the big ideas came first,
string theory emerged, originally in the early 1970's, by a process
of tinkering without anyone having the big picture.  At first, one could
see only the loftiest peaks above the clouds.  For thirty years now,
we have chipped away, uncovering some of the underlying structure,
but with much still unclear.  

One important development in the 1990's was that, as long as one can 
ignore supersymmetry breaking, we learned to extrapolate to large
values of parameters such as $e^2/4\pi \hbar c=e^\phi$.  We always could
calculate for $e^2/4\pi \hbar c<<1$; now we can say something for $e^2/4\pi \hbar c
>>1$.  

The basic technique is to study magnetic monopoles and other nonperturbative
excitations.  For $e^2/4\pi \hbar c$ the monopoles are extremely heavy compared
to the ordinary particles, and this makes them less important.  However,
for $e^2/4\pi \hbar c$ large, the monopoles become light and we should find a 
description of the theory in terms of monopoles.  

At the cost of oversimplifying things a bit, I will give a qualitative
explanation of this.  According to Dirac, the magnetic charge of a 
magnetic monopole is $g=2\pi \hbar c/e$.  This shows that $g$ is much
larger than $e$ when $e$ is small, and much smaller than $e$ when $e$ is
large.  So to the extent that the mass of an electron or a magnetic
monopole has an electromagnetic origin, one would expect that monopoles
are the light objects for $e\to\infty$.

So for $e^2/4\pi \hbar c\to \infty$, we need to use a description that
favors monopoles rather than electrons, but this turns out to be
another string theory, or a close cousin.  Thinking along these lines
led to the understanding that there is only one string theory: the five
string theories as traditionally understood are different limiting cases
of one more complete theory, sometimes called $M$-theory.  $M$-theory is
the candidate for superunification, though we still do not really
understand it.

\bigskip\noindent{\it Magnetic Monopoles}

Monopoles have played an important role in developing this picture,
so in particular, if string theory is correct, they should exist. Where are they?  Inflation may
have made the monopole density unobservably small; this was plausible
when first proposed over 20 years ago, and remains plausible.  But it would
be a shame, so let us hope that a reasonable number of monopoles survived
somehow.

What is the monopole mass?  The standard picture based on GUT-like
models leads to monopole masses above the GUT scale, say in the range
of $10^{17}$ to $10^{20}$ GeV.  (The upper bound is the mass of
a Reissner-Nordstrom black hole carrying magnetic charge; the lightest
monopole could scarcely be heavier than this.)

Monopoles of such high mass are not affected by the Parker bound on
monopole flux, because in a galaxy the gravitational force on such a monopole
is much greater than the magnetic force.  As far as I know, they would
have been viable dark matter candidates, except that bounds from MACRO
\cite{macro}
and other experiments exclude the hypothesis that monoples, even at the
highest plausible mass, are the dominant component of the galactic halo.

Along with GUT-like models, we also should consider models in which
a low scale unification with gravity is achieved, via large extra
dimensions \cite{dimop} or a warped scenario \cite{sundrum}.  For example, if
 such unification occurs
at energies of a few TeV, we should expect monopoles at perhaps 
10 - 100 TeV.  Such light monopoles would be accelerated in the galactic
magnetic field to ultrarelativistic energies, so their experimental
signatures are completely different from those of the GUT-scale monopoles.
(For a recent assessment, see \cite{wkwb}.)
Also, they might not have been diluted by inflation.  It is hard to 
be specific on this last point, as current ideas for low-scale unification
are really scenarios more than detailed models.  

Another interesting property of monopoles in string models \cite{wenwit} is that
in many models, the minimum magnetic charge is not the Dirac quantum
$2\pi \hbar c/e$, but is larger than this by an integer $n$ that depends on
the model.  Reciprocally, these models have unconfined massive particles
with electric charge $e/n$.  These particles would have GUT-like masses
in the GUT-scale models and  would be at the TeV energy scale in models
with low energy unification.  The cosmic ray flux of particles with electric
charge equal to or greater than $e/5$ and $\beta>1/4$ has been significantly constrained
by MACRO \cite{moremacro}.

Ironically, though models with TeV scale unification with gravity
are sometimes seen as making TeV scale supersymmetry unnecessary,
they also make it inevitable, if string theory is correct.  
To see this, let us perform a simple thought experiment (which
we  could have considered earlier in the case of the monopoles).
Suppose that accelerator experiments at TeV energies do discover a
unification with quantum gravity and an effective higher-dimensional
Planck scale of a few TeV.  Then, if it is true (as string theory
seems to tell us) that quantum gravity requires supersymmetry at the 
fundamental Planck scale, we should expect to find supersymmetry at
energies of a few TeV, if not less!

A nice illustration of this last point is in a recent paper by
Giddings and DeWolfe \cite{giddings}.  (See also \cite{luty} for a related discussion.)
They consider a warped scenario with an underlying ten-dimensional
Planck mass that for most of the states is effectively at or above the usual
GUT scale; moreover, supersymmetry is broken at this high scale.  However,
because of the warping of the metric, some of the states survive
down to TeV energies and below.   In this model, the TeV scale 
particles include some excited gravitons, and by experiments at TeV
energies involving this low energy sector (which is assumed to contain
the familiar particles), one could observe unification of elementary
particles with gravity.  And sure enough, as our heuristic argument
would predict, Giddings and DeWolfe find that although the GUT-scale
particles have GUT-scale supersymmetric mass splittings, the TeV
scale particles have TeV-scale supersymmetric splittings.  Thus,
experiments that uncover unification with gravity would also have the energy to uncover
supersymmetry.

So in short, weak scale monopoles and weak scale supersymmetry breaking
should be general consequences of models with weak scale unification
with gravity.  

Where else might we see signatures of string theory?  It would be
nice to detect some of the numerous stringy scalar fields either because of
changes in time of natural constants (because of evolution of scalar
fields in the vacuum) or because of departures from the equivalence 
principle, or deviations from Newton's law of gravity at short distances.
Concerning the changes in natural constants, I am personally a little
pessimistic because I suspect that scalars whose evolution causes
a detectable change in natural constants might have already shown
up in tests of the equivalence principle.\footnote{To spell this out a bit,
 scalars that are evolving so slowly as to produce a change
in natural constants on cosmic time scales are so light that their
Compton wavelengths are far greater than any length scale that is relevant
in tests of
the equivalence principle or general relativity.  Such particles
would therefore be detected in such experiments unless their couplings
are significantly weaker than gravitational.
By contrast, since
current laboratory tests of gravity involve shorter distances than those
explored in the other experiments, a scalar of the right mass might
produce a detectable departure from Newton's laws of gravity without showing
up in tests of the equivalence principle or general relativity.}
But let's hope.

Among various other conceivable possibilities, I will only add that
the very name  ``string theory'' suggests that if this
theory is right, strings may exist in the heavens.  (Again, the cosmic strings
may be greatly diluted by inflation; this remark applies especially
to heavy ones.)  One can imagine elementary
strings stretching across the sky, and other conceivable objects with
GUT-like tension, possibly visible in the CMB maps.  But also
plausible are other, lighter objects, perhaps all the way down to
the TeV scale (which corresponds to a string tension of roughly
$10^{-4}$ ${\rm gm/cm}$).  
The light objects might arise, for example, from spontaneous breaking
of extra $U(1)$ factors of the gauge group  that we do not
yet know about.  The light ones would not be detectable gravitationally,
but they might be detected if they are abundant in the Milky Way
(or the Solar System? \cite{vilen}), in which case, if they are superconducting
(not unlikely for strings that arise from breaking of extra $U(1)$'s
\cite{witten}), they might be detectable by causing magnetic disturbances
and creating antimatter clouds when they interact with a magnetic field.


\end{document}